\documentclass[prb,preprint,amsmath,amssymb]{revtex4}
\begin{document}
\author{Kevin Leung}
\affiliation{Sandia National Laboratories, MS 1415, Albuquerque, NM 87185\\
\tt kleung@sandia.gov (505)8441588}
\date{\today}
\title{Predicting the Voltage Dependence of Interfacial Electrochemical
Processes at Lithium-Intercalated Graphite Edge Planes}

\input epsf
%\ssp

\begin{abstract}
 
The applied potential governs lithium-intercalation and electrode passivation
reactions in lithium ion batteries, but are challenging to calibrate in
condensed phase DFT calculations.  In this work, the ``anode potential''
of charge-neutral lithium-intercalated graphite (LiC$_6$) with
oxidized edge planes is computed as a function of Li-content ($n_{\rm Li}$)
at edge planes, using {\it ab initio} molecular dynamics (AIMD), a
previously introduced Li$^+$ transfer free energy method, and the experimental
Li$^+$/Li(s) value as reference.  The voltage assignments are corroborated
using explicit electron transfer from fluoroethylene carbonate radical anion
markers.  PF$_6^-$ is shown to decompose electrochemically (i.e., not just
thermally) at low potentials imposed by our voltage calibration technique.
We demonstrate that excess electrons reside in localized states-in-the-gap
in the organic carbonate liquid region, which is not semiconductor-like
(band-state-like) as widely assumed in the literature.

\vspace*{0.5in}
\noindent keywords: lithium ion batteries; voltage prediction;
density functional theory; {\it ab initio} molecule dynamics;
computational electrochemistry

\end{abstract}

\maketitle

\section{Introduction}

The applied potential governs the thermodynamics and kinetics of lithium
ion battery (LIB) interfacial processes.  Li$^+$ insertion into graphite
anodes to form LiC$_6$ is completed at 0.1~V vs.~Li$^+$/Li(s),
or $\sim$$-2.9$~V vs.~standard hydrogen potential.  Before this low voltage
is reached, commonly used battery electrolytes containing ethylene carbonate
(EC), cosolvents, and Li$^+$/PF$_6^-$ salt already decompose 
at 0.7-0.8~V vs.~Li$^+$/Li(s).  Fortunately, the growth of self-limiting
films (called ``solid-electrolyte interphase,'' or SEI) formed via
electron-injection-induced sacrificial electrolyte degradation passivates 
and stabilizes the anode.\cite{review0,review}  SEI considerations are
also relevant for new anode materials like silicon.  To intepret measurements
and to help devise better artificial SEI/passivation layers,\cite{dillon}
there is a need to use electronic structure computational tools (e.g.,
Density Functional Theory, DFT) to predict the voltage dependence of
liquid-solid interfacial processes.  Such a capability will have significant
impact for studying not just LIB,\cite{tateyama,greeley} but also
lithium-air batteries,\cite{bryantsev} water-splitting
processes,\cite{selloni,sprik12} and broad areas relevant to fuel cells,
catalysis,\cite{otani08,casewestern,rossmeisl13,gross} and
electrodeposition.\cite{schmickler,zavadil} In this work, we validate a
recently devised potential calibration scheme,\cite{voltage} apply it to
interfaces between liquid EC and oxidized edge planes of LiC$_6$, and explore
the possibility of electrochemical decomposition of the counter-ion (PF$_6^-$)
used in commercial LIB electrolytes.  In the electronic supporting
information (S.I.), we document the significant error that can arise if the
liquid electolyte is omitted in voltage estimates at interfaces.

For non-redox active systems like non-Faradaic supercapacitors, the voltage
difference between two electrodes arises from their different surface charges
mediated by electric double layers in the liquid regino.\cite{borodin}  In
contrast, on complex LIB electrode surfaces, what a certain ``applied voltage''
means at the atomic level has not been sufficiently conceptualized, partly due
to the difficulty in probing details at such lengthscales.\cite{harris}  On
the theory side, potential calibration has been challenging in periodic
boundary conditions, condensed-phase DFT simulations that depict liquid-solid
interfaces.\cite{sprik12,otani08,casewestern,rossmeisl13,gross}  
DFT calculations are performed at constant number of electrons, not constant
voltage.  Each DFT simulation cell is associated with one electrode/Fermi
level and is incompatible with a second reference electrode.  Furthermore, most
DFT electrochemistry calculations are conducted at T=0~K, which precludes
explicit treatment of liquid solvents, dissolved salts, and, in the majority of
cases, charged interfaces.  Several interesting recent advances have coupled
the Poisson-Boltzmann equation or related approaches to DFT at the expense of
introducing vacuum-solid or -liquid interfaces into the
model.\cite{otani08,casewestern}  (The vacuum layer in effect serves as the
reference electrode.)  Purely condensed phase simulations with no vacuum
region require other methods.

We recently estimated what will be called the ``anode potential''
(${\cal V}$) of inert LiC$_6$ basal planes at finite temperature.\cite{voltage}
The justification is briefly and heuristically described here.  Our approach
seeks to mimic LIB experimental processes, where Li-deintercalation occurs
via transfer of Li$^+$ from LiC$_6$, through the liquid-solid interface, to
the liquid electrolyte, and ultimately into the Li metal counter electrode
not explicitly depicted in the simulation.  $e^-$ flows in the same direction,
but through the external circuit.  Experimentally, it is known that these
charge transfer processes occur at the onset potential of 0.1~V
vs.~Li$^+$/Li(s).

We model this half-cell reaction at the onset of LiC$_6$ delithiation,
\begin{equation}
({\rm LiC}_6)_n \rightarrow ({\rm Li}_{(1-1/n)}{\rm C}_6)_n^- 
	+ {\rm Li}^+({\rm solv}).  \label{eq1}
\end{equation}
While our model does not include counter electrodes, and the excess $e^-$
is left on the anode in the simulation (with finite surface area correction,
see below), Eq.~\ref{eq1} has effectively completed the $e^-$ circuit.  This
is because, at equilibrium and in the absence of load-induced voltage
drop, the Fermi level ($E_{\rm F}$) of the Li metal ``counter electrode'' 
in our thought experiment must be lowered by 0.1~V to coincide with the
$E_{\rm F}$ of LiC$_6$.  Under these conditions, the excess $e^-$ on
Li$_{1-\delta}$C$_6$ can start to flow to Li(s).  By reaching equilibrium
(tuning the free energy change ($\Delta G_t$) of the Eq.~\ref{eq1} to zero)
via varying the surface charge ($\sigma$) on the electrode surface, we arrive
at the experimentally known half-cell voltage for this reaction.  This is the
reference point that can be pegged to measurements.  Away from this
${\cal V}$=0.1~V vs.~Li$^+$/Li(s) fixed point, anode voltages are clearly
related to the free energy of monovalent Li$^+$ transfer between LiC$_6$ and
EC liquid ($\Delta G_t$), $\Delta {\cal V}=-\Delta G_t/|e|$, provided that the
interior Li atoms are frozen and not allowed to leak into the electrolyte
(i.e., there is no redox reaction), which is the case in our
simulations.\cite{voltage} The $\sigma$-${\cal V}$ relationship associated
with our frozen-Li basal plane LiC$_6$ model electrode is reminiscent of
those in non-Faradaic supercapacitors.

The {\it ab initio} molecular dynamics (AIMD, or DFT/MD) technique is
used to calculate $\Delta G_t$.\cite{voltage}  $\Delta G_t$ calculations 
deal with physical ions, not infinitesimal/theoretical test charges, and
therefore circumvent formal/unmeasurable concepts like the difference between
``Volta'' and ``Galvani'' definitions of the potential.\cite{sprik12,pratt92} 
Further justification, including a thought experiment on explicitly
including the interface between the liquid electrolyte and the Li metal
reference electrode, and comparison between our approach and 
related methods found in the aqueous computational electrochemistry
literature, are given in Ref.~\onlinecite{voltage}.

The present work focuses on LiC$_6$ edge planes through which Li$^+$ can
intercalate and deintercalate.  Edge planes are far more technologically
relevant and complex than the proof-of-principle pristine basal plane
considered previously.\cite{voltage}  But the same theoretical method can be
used to examine the edge plane voltage.  Indeed, Ref.~\onlinecite{voltage}
(Fig.~2) implies that this Li$^+$ transfer protocol can in priciple directly
compare the voltage on any two electrodes.  This is because a sufficiently
thick liquid electrolyte region intervening between a basal and an edge
plane can chemically and electrostatically screen them from each other.  So
the difference in equilibrium free energies of Li$^+$ transfer from each
electrodes to the electrolyte, divided by $|e|$ should be proportional 
to their voltage difference.  Since LiC$_6$ is an electronic conductor,
at equilibrium $e^-$ flows between surfaces to make the voltage the
same on basal and edge planes.  On pristine edge planes, the all-important
$\Delta G_t$=0 point is determined not only by the electronic charge
($\sigma$) compensated by mobile Li$^+$ in the liquid electrolyte, but also
the fraction of Li occupying the edge sites ($n_{\rm Li}$).  At a fixed,
applied potential, these two quantities should adjust themselves to minimize
the free energy of the system.  For the illustrative purpose of this
work, we have fixed $\sigma$=0 and only varied $n_{\rm Li}$, and ${\cal V}$
can be considered the instantaneous voltage before the edge Li$^+$ content
can change. 

Our predicted edge plane voltage as $n_{\rm Li}$ varies will be corroborated
using explicit electron transfer from fluoroethylene carbonate radical anion
markers.  Using AIMD simulations with appropriately calibrated potentials,
PF$_6^-$ is shown to exhibit concerted $e^-$ transfer and bond-breaking
reactions at low voltages, suggesting that under such conditions electrochemical
decomposition may need to be considered.  This is significant because it is
widely accepted that PF$_6^-$ decomposes only thermally or due to reaction
with trace water.\cite{kostecki_pf6,plak_pf6}  Finally, the simulation cells
used in this work provide information about the electronic orbital
alignment at explicit electrode/electrolyte interfaces.  We show that excess
electrons reside in localized states-in-the-gap in the organic carbonate
liquid region, not in delocalized states at the conduction band minimum.  The
band structure is {\it not} semiconductor-like (band-state-like) as widely
assumed in the literature.\cite{goodenough}

\section{Method}

Our simplified electrode model consists of a LiC$_6$ strip with all C=O
termination.\cite{ec,mccreery}  Four neighboring C=O bonds form a pocket
where Li$^+$ can reside. The Li$^+$ surface density ($n_{\rm Li}$) is unity
if all such pockets are occupied.  Figure~\ref{fig1} depicts the periodically
replicated 29.74$\times$14.97$\times$15.06~\AA$^3$
interfacial simulation cell containing a Li$_x$C$_{192}$O$_{48}$ anode slab
and 32~EC molecules.  Interior atoms in the anode, with LiC$_6$ stochiometry,
are frozen while C, O, and Li atoms at the edges are allowed to move.
$n_{\rm Li}<1$ electrode configurations are obtained by randomly, and as
uniformly as possible, removing Li atoms from the two LiC$_6$ edge planes.  

\begin{table}\centering
\begin{tabular}{  l c c c r r r  } \hline
 & $n_{\rm Li}$ & $N$(Li) & $\lambda$ & $t_{\rm max}$ &
$\langle dH(\lambda)/d\lambda \rangle_\lambda$ & $\Delta G_t$ \\ \hline
A& 0.417 & 0 & 0.211 & 43.9 & +5.45$\pm$0.11 & \\
B& 0.417 & 0 & 0.789 & 44.4 & -6.63$\pm$0.04 & -1.14 \\ \hline
C& 0.500 & 0 & 0.211 & 52.1 & +5.86$\pm$0.10 & \\ 
D& 0.500 & 0 & 0.789 & 40.8 & -6.31$\pm$0.10 & -0.78 \\ \hline 
E& 0.583 & 0 & 0.211 & 26.1 & +6.59$\pm$0.11 & \\
F& 0.583 & 0 & 0.789 & 20.9 & -6.23$\pm$0.11 & -0.36 \\ \hline 
G& 0.417 & 1 & 0.211 & 25.0 & +5.80$\pm$0.05 & \\
H& 0.417 & 1 & 0.789 & 25.2 & -6.61$\pm$0.07 & -0.94 \\ \hline
%I& 0.417 & 2 & 0.211 & 13.5 & +5.72$\pm$0.06 & \\
%J& 0.417 & 2 & 0.789 & 13.4 & -6.32$\pm$0.03 & -0.84  \\ \hline 
\end{tabular}
\caption[]
{\label{table1} \noindent
Details of AIMD trajectories for $\Delta G_t$ calculations.  $N$(Li) is 
the number of mobile Li$^+$ in the liquid region, and $\lambda$ is the
net charge of the Li$^{\lambda +}$ ion frozen in the middle of the liquid.
$t_{\rm max}$ is the total trajectory duration in picoseconds, and include
the first 1~ps equilibration discarded when collecting statistics.  The
exceptions are C \&~D, where $t_{\rm max}$ includes the discarded first
10~ps.  Integrands and $\Delta G_t$ are in eV; the latter is obtained by
averaging the two integrands, and includes a $-0.39$~eV entropic correction
and a $-0.15$~eV correction for using a 2-point treatment of Li$^+$
solvation.\cite{voltage} To convert ($-\Delta G_t$) to ${\cal V}$, add
0.1~V for Li$^+$/Li(s) and 0.1~V for the ``half $e^-$ rule" (see text).
}
\end{table}

AIMD trajectories and a two-point
thermodynamic integration (T.I.) formula with corrections/extrapolations are
used to compute $\Delta G_t$ (Table~\ref{table1}).  These simulations apply
the VASP code\cite{vasp} with PAW pseudopotentials\cite{paw} and the DFT/PBE
functional.\cite{pbe}  An energy cut-off of 400~eV, 10$^{-6}$~eV wavefunction
convergence, and $\Gamma$-point sampling are enforced.  Spot checks show
that 1$\times$2$\times$2 $k$-point sampling changes the integrands in
$\Delta G_t$ by less than $\sim$0.05~eV, similar to basal plane
cases.\cite{voltage}  A thermostat keeps the trajectories at an elevated
T=450~K to improve statistics and prevent EC crystallization.  Tritium masses
are substituted for proton masses to permit 1~fs time steps.  Compared to
the previous work,\cite{voltage} the predicted $\Delta G_t$ is shifted by
$-0.17$~V to correct the prior neglect of quantum nuclear effect inside
bulk LiC$_6$ and the inadvertent use of T=450~K when adding translational
and vibrational entropies to compare with experiments performed at T=300~K.  

The trajectories are initiated using Monte Carlo (MC) simulations in which
anode atoms are frozen in DFT/PBE-optimized configurations.  LIB salt
concentration is typically 1.0~M, the static dielectric constant is large
(Debye length $\sim$3~\AA), and electrode surfaces should be screened
from each other even in a small simulation cell.  Electrical double
layers should be well-equilibrated to the extent that the simple classical
force fields used are accurate.  More MC details are described in the S.I.

\section{Results}
\subsection{Controlling Potential at LiC$_6$ Edge Planes}

\begin{figure}
\centerline{\hbox{ (a) \epsfxsize=3.00in \epsfbox{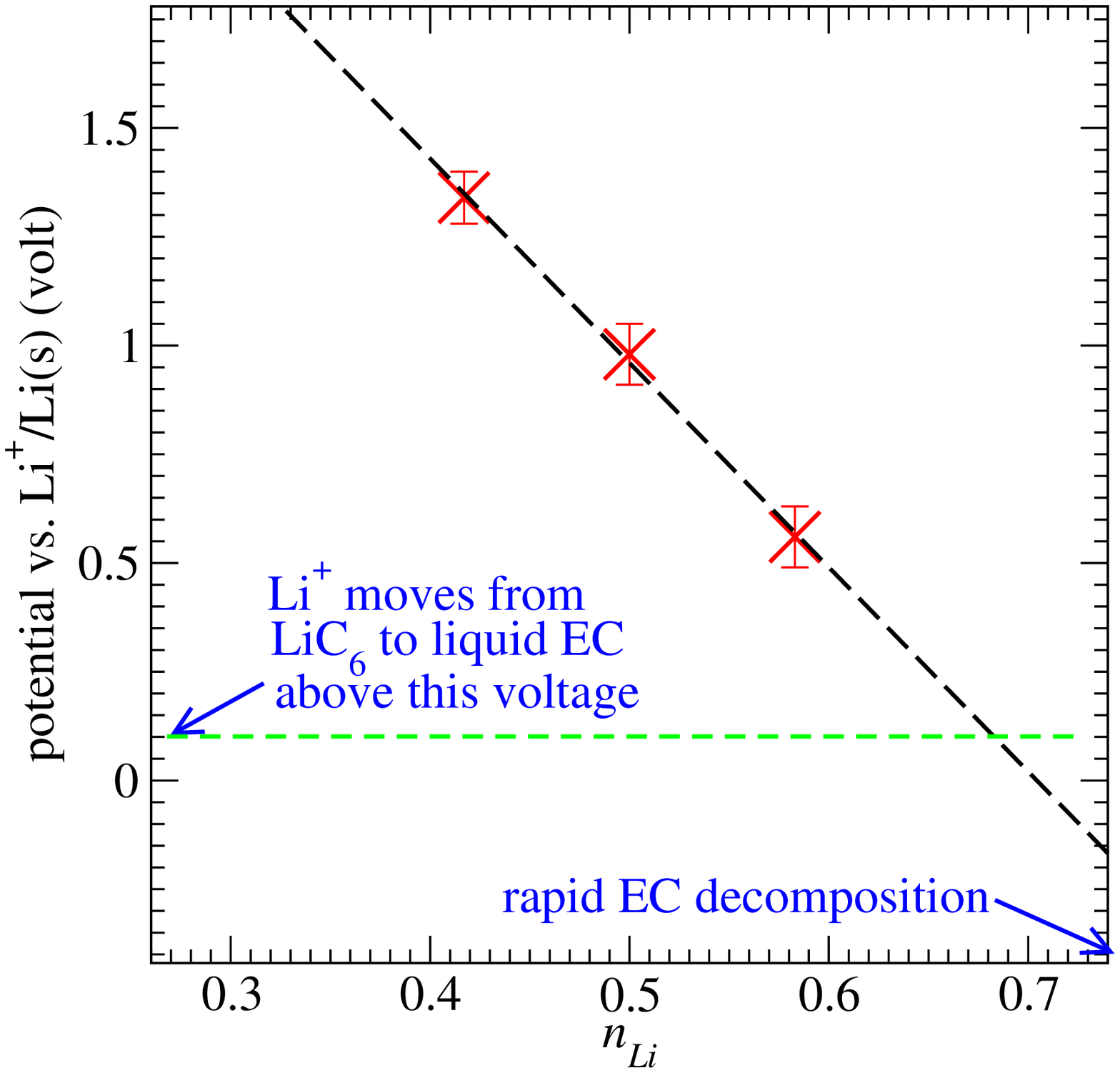} }}
\centerline{\hbox{ (b) \epsfxsize=3.00in \epsfbox{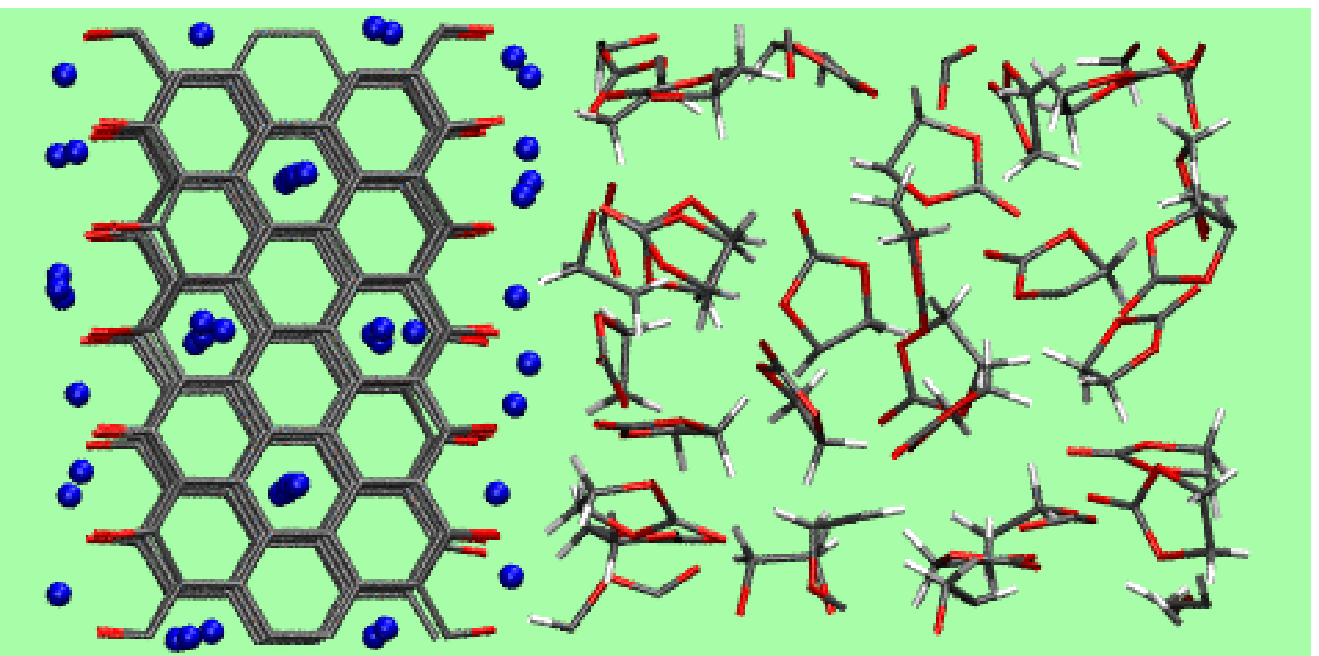} }}
\caption[]
{\label{fig1} \noindent
(a) ${\cal V}$ at zero electronic charge as a function of edge plane Li
content.  The dashed line is a linear extrapolation.  (b) A snapshot of
the edge plane interfacial simulation cell.  C, O, H, and Li are grey,
red, white, and blue, respectively.
}
\end{figure}

Instead of mapping the entire two-dimensional potential 
${\cal V}(n_{\rm Li},\sigma)$, we focus on $\sigma=0$.  In
Fig.~\ref{fig1}a, the linearly extrapolated ${\cal V}(\sigma=0,n_{\rm Li})$
reaches the LiC$_6$ experimental plateau voltage of 0.1~V vs.~Li$^+$/Li(s) at
$n_{\rm Li}$$\sim$0.69.  If $n_{\rm Li}$$>$0.69, $\sigma >0$ would be needed
to raise ${\cal V}(\sigma,n_{\rm Li})$ back to the green line and achieve
the experimental potential associated with LiC$_6$.  This merely means that
some of the edge Li must then be considered Li$^+$ ions --- not atoms ---
compensated with mobile PF$_6^-$ further away in the electrolyte.\cite{conceit}

Note that a ``half-electron rule'' vertical shift has been included to convert
$\Delta G_t$ (Table~\ref{table1}) to ${\cal V}(\sigma=0,n_{\rm Li})$
(Fig.~\ref{fig1}a).  T.I.~calculations involve moving a Li$^+$ from LiC$_6$ to
the middle of the solvent region of a charge-neutral simulation cell, leaving
an $e^-$ behind.  Along the T.I.~path, an average of half an $e^-$, or
$\sigma$=$-|e|$/($4A)$, exists on the two electrode surfaces, where $A$ is
the lateral surface area.  We assume this excess charge is uniformly
distributed on the conducting electrode surfaces, in accordance with classical
electrostatic predictions; see the S.I. of Ref.~\onlinecite{voltage} for 
analysis of instantaneous charge distributions.  The effect of this average
$\sigma$ is estimated by finite difference and subtracted from
Fig.~\ref{fig1}a, as follows.  We inject one mobile Li$^+$, compensating an
excess $e^-$ on the electrode, and recompute $\Delta G_t$ (Table~\ref{table1},
trajectories G-H).  The simulation cells remain charge-neutral.  
$\Delta {\cal V}$ is found to be $-0.20$~V ($\delta \sigma/ \delta
{\cal V}$=$17.8$~$\mu$C/(cm$^2 V)$) after adding the one $e^-$.  This is
smaller in magnitude than that in basal plane simulation cells.\cite{voltage}
To undo this surface charging effect, a $+0.1$~V correction is thus applied.
The shift vanishes at large $A$, and represents an extrapolation to infinite
system size.  We stress that experimentalists can impose a potential without
knowing details about the surfaces, but DFT calculations work differently;
$\sigma$ and $n_{\rm Li}$ need to be adjusted to arrive at the desired voltage.

In the absence of the liquid electrolyte, the potential at zero surface
charge is directly related to the work function of the electrode in vacuum.
In the S.I., we report work functions and show that the potential in vacuum,
predicted as a function of $n_{\rm Li}$, is significantly modified by the
inclusion of the liquid electrolyte in the main text.  This observation
dovetails with predictions that work functions of metals can vary by $\sim$1~V
when their surfaces are covered with a monolayer of water\cite{gross} or
organic molecules.\cite{brocks}

Although linear extrapolation is not expected to hold over the entire
$n_{\rm Li}$ range, ${\cal V}(n_{\rm Li},\sigma=0)$ appears to extrapolate
to a value substantially below 0~V vs.~Li$^+$/Li(s) at $n_{\rm Li}$=1.
Negative potentials relative to Li$^+$/Li(s) are below the operating conditions
of LIB anodes.  This strongly suggests that $\sigma$=0 and $n_{\rm Li}$=1
yield an overpotential for electrolyte decomposition.  In the literature, C=O
edge AIMD simulations have been reported at $n_{\rm Li}$=1 and $\sigma$=0, and
EC molecules are found to decompose in picosecond via two different 2-$e^-$
mechanisms, releasing CO and C$_2$H$_4$ gases, respectively.\cite{ec}  What
are the potential dependences of these two competing processes?  Recently,
it has been predicted that the 2-$e^-$ CO-releasing route has a far lower
reaction barrier than C$_2$H$_4$ generation in bulk liquid electrolyte
regions.\cite{cpl}  It is therefore unlikely that C$_2$H$_4$ should be
a dominant 2-$e^-$ product unless there is an overpotential.  Consistent
with this deduction, Fig.~\ref{fig1} indeed suggests the Ref.~\onlinecite{ec}
system, where $n_{\rm Li}=1$, is at ${\cal V}<0$~V.  If this were not the
case, mostly CO products are expected.  Note that two-electron reduction can
yield C$_2$H$_4$ gas if $e^-$ are added sequentially, not simultaneously,
separated by milliseconds.\cite{cpl}  This timescale is far larger than
our AIMD trajectory durations.

\subsection{Validating Predicted Potential: Electron Transfer}

\begin{figure}
\centerline{\hbox{ \epsfxsize=3.00in \epsfbox{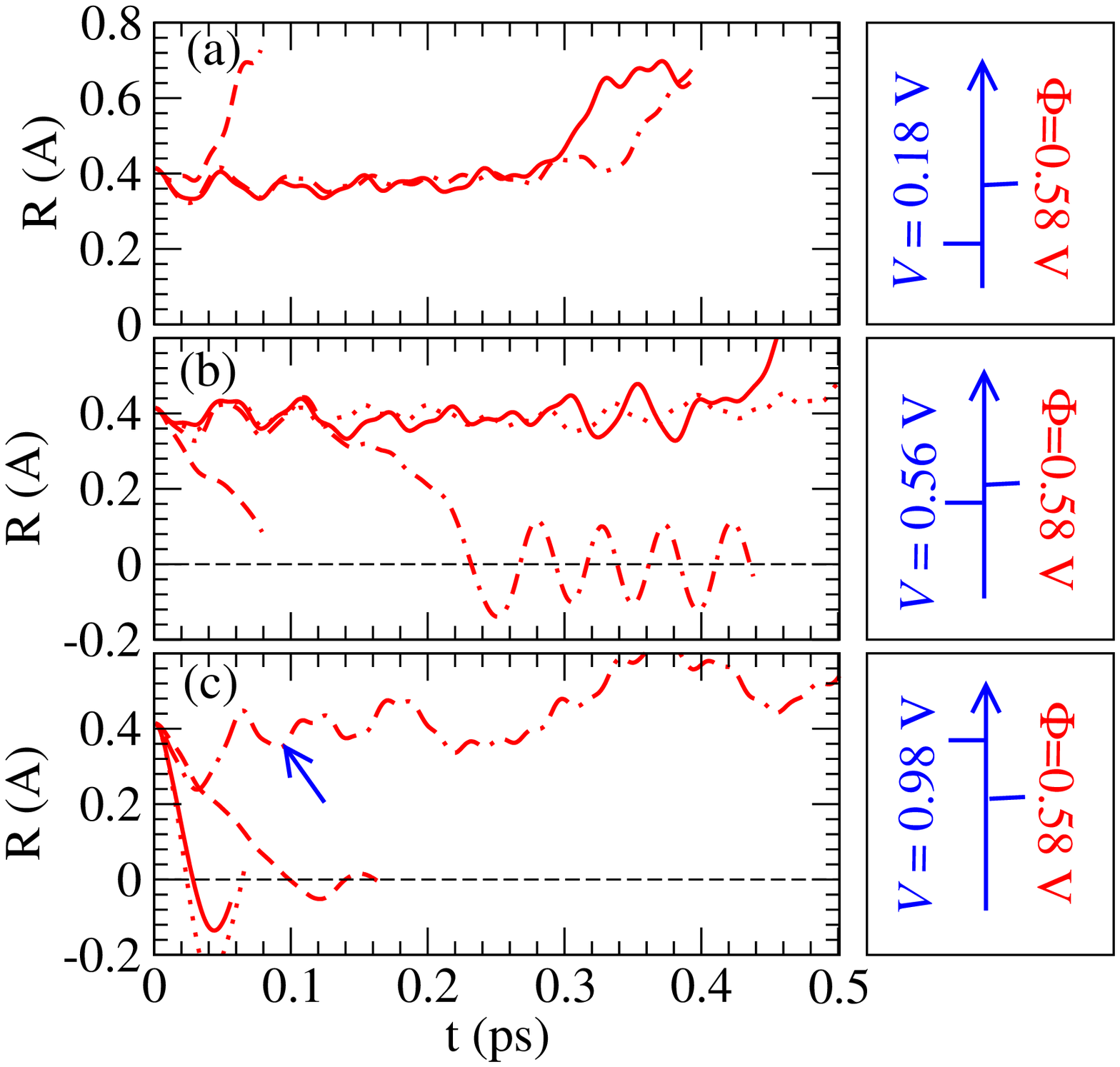} }}
\centerline{\hbox{ (d) \epsfxsize=1.50in \epsfbox{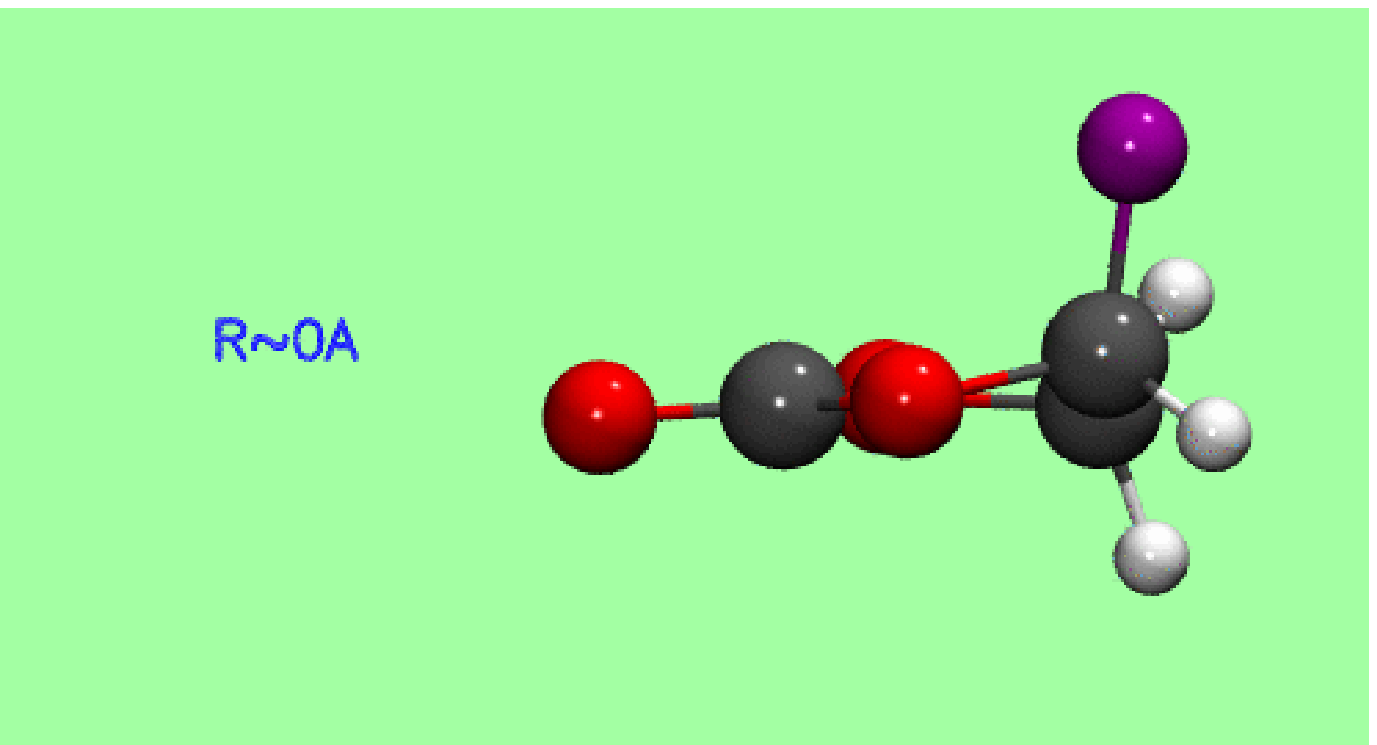} }
            \hbox{ (e) \epsfxsize=1.50in \epsfbox{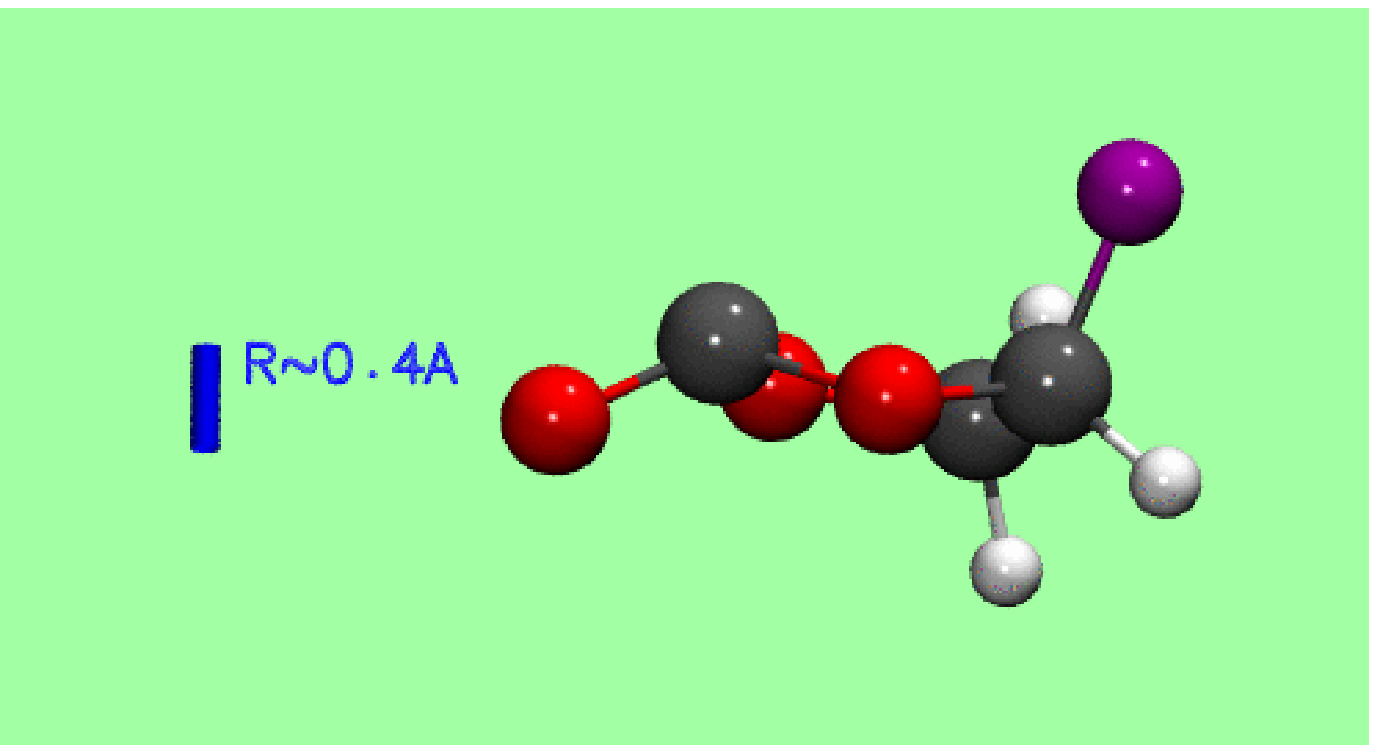} }}
\caption[]
{\label{fig2} \noindent
(a)-(c) $R$ as a function of time at various anode potentials ${\cal V}$.
Different line shapes denote trajectories with different initial
configurations.  (d)\&(e) Optimized FEC and FEC$^-$ molecules.  Out-of-plane
coordinates $R$ are illustrated.  Purple circles represent F~atoms.
}
\end{figure}

The predicted potential should not depend on whether Li$^+$ or $e^-$ moves
across the interface. Next we demonstrate that ${\cal V}(\sigma=0,n_{\rm Li})$,
calibrated using Li$^+$ transfer above, is also consistent with $e^-$ transfer.
In the middle of the liquid region is placed a fluoroethylene carbonate
(FEC$^-$) radical anion (Fig.~\ref{fig2}), which is an effective electrolyte
additive molecule for improving SEI on anode surfaces.\cite{fec}  In
charge-neutral FEC (Fig.~\ref{fig2}c), the carbonyl carbon (C$_{\rm C}$) is
coplanar with the three O~atoms.  In contrast, FEC$^-$ is bent
(Fig.~\ref{fig2}d), with C$_{\rm C}$ now $sp^3$ hybridized.  This leads to a
large ``reorganization energy'' in the Marcus theory sense,\cite{marcus}
discussed below.  We define the C$_{\rm C}$ out-of-plane displacement $R$
as the scalar product between (i) the normalized vector product connecting
the three O-atoms, and (ii) $({\bf R_{\rm C}}-{\bf R_{\rm Oave}})$, where
``Oave'' is the mean position of the three oxygen atoms.  
$R$=0.0~\AA\,~and~0.4~\AA\, in the optimized FEC and FEC$^-$ geometries.
Monitoring $R$ grants easy access to FEC charge states.  To initiate AIMD
simulations, we first conduct MC using classical force fields (see the S.I.),
then run AIMD for 1.0~ps while freezing all FEC$^-$ atoms, and finally remove
FEC constraints at ``$t$=0'' of AIMD trajectories.  The time evolution of
$R$ are shown in Fig.~\ref{fig2}.

The reduction potential of FEC in bulk liquid regions is predicted to be
$\Phi$=0.58~eV when it is not coordinated to Li$^+$ (S.I.).  If
${\cal V}(\sigma=0,n_{\rm Li})$$<$$\Phi$, the excess $e^-$ should stay on
FEC$^-$, ultimately leading to a second reduction of FEC$^-$ and rapid
decomposition.\cite{fec,note3}  In the opposite case, the excess $e^-$
should be transferred to the electrode.  The S.I.~shows that this expectation
is always satisfied with different initial configurations on less
electrochemically active basal plane surfaces.  At edge planes, this test is
successful, but less than 100~\% of the time.  When ${\cal V}$$\sim$0.18~V
($n_{\rm Li}$=0.667), lower than $\Phi$=0.58~V, FEC$^-$ persists in all
trials until it absorbs a second $e^-$ and decomposes (Fig.~\ref{fig2}a).
This dovetails with our expectation.  When ${\cal V}$=0.56~V, very close to
$\Phi$, FEC$^-$ is stable for hundreds of femtoseconds until it decomposes in
two out of four trials; in the remaining two cases FEC$^-$ loses its electron
(Fig.~\ref{fig2}b).  Given our statistical uncertainties, this 50/50 split
in the outcome is reasonable.  When $\Phi$$<$${\cal V}$=0.98~V, FEC$^-$
should give up its excess $e^-$.  This is observed 3 out of 4 times
(Fig.~\ref{fig2}c).

On the whole, Fig.~\ref{fig2} demonstrates that ${\cal V}$ calibrated using
Li$^+$ transfer also correctly governs $e^-$ transfer.  The one glaring
``error'' at ${\cal V}$=0.98~V is apparently due to overly rapid C-O bond
breaking in FEC$^-$ predicted using the PBE functional.  This occurs within
100~fs (arrow in Fig.~\ref{fig2}c; the predicted MP2 barriers for breaking
this bond in EC$^-$ and FEC$^-$ are consistent with slower reaction
rates.\cite{cpl,fec}).  Afterwards, the FEC$^-$ ring cannot be reformed in
AIMD timescale even if the anode potential favors it.   It is also worth
pointing out that DFT/PBE allows unphysical splitting of an excess $e^-$ 
between the electrode and the redox center, artificially accelerating $e^-$
transfer rate,\cite{wtyang} and may make Fig.~\ref{fig2} unduly sensitive to
initial configurations.  Instantaneous fluctuations in the potential
experienced by redox centers are physical and real, but $e^-$ transfer occurs
over a finite timescale that partially averages out the fluctuations.  Despite
this, the overall trend of ${\cal V}$ and $\Phi$ correspondence is correctly
predicted: as ${\cal V}$ decreases, FEC$^-$ retains its excess $e^-$ more
readily.

\subsection{Excess Electrons Form Localized States in the Gap}

\begin{figure}
\centerline{\hbox{ \epsfxsize=3.00in \epsfbox{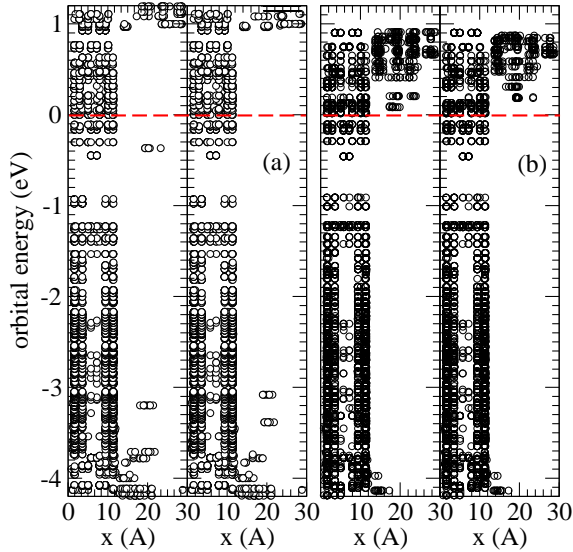} }}
\caption[]
{\label{fig3} \noindent
Instantaneous Kohn-Sham orbital decomposed on to atoms at their
$x$-coordinates at $t$=0~ps in Fig.~\ref{fig2}b\& c.  $0<x<13.4$~\AA\,
denotes the metallic LiC$_6$ region; outside that range resides the liquid
electrolyte which has a wide band gap.  The Fermi level is at $E$=0.0~eV.
Panels a~\&~b depict the majority and minority spin channels of on trajectory
each in Fig.~\ref{fig2}a and Fig.~\ref{fig2}b.  The excess $e^-$ ``polaron'' 
orbital of FEC resides in the majority spin channel, near $E_{\rm F}$, in 
the electrolyte region.
}
\end{figure}

Figure~\ref{fig3} depicts spatially-demarcated DFT Kohn Sham spin-orbital
levels at $t$=0~ps in two trajectories taken from Fig.~\ref{fig2}b~\&~c.
In panel (a) (${\cal V}$=0.56~V), the highest occupied molecular orbital
(HOMO) is localized on FEC$^-$ --- reminiscent of a polaron in solid state
physics --- just below the Fermi level ($E_{\rm F}$).  FEC$^-$ is found to
decompose from this configuration.  In panel (b), $V$=0.98~V, higher than
$\Phi$=0.58~V; the localized orbital resides above $E_{\rm F}$, and FEC$^-$
loses its excess $e^-$ rapidly in this trajectory.  Fig.~\ref{fig3}
illustrates that, for organic carbonate-based liquid electrolytes, the 
thermodynamic onset of electrolyte electrochemical reduction and SEI formation
occurs when $E_{\rm F}$ coincides with the organic carbonate {\it localized}
orbital found in the gap between liquid HOMO and liquid LUMO (lowest
unoccupied molecular obrital or liquid conduction band minimum).\cite{note4}
SEI formation does {\it not} begin, thermodynamically speaking, when the
electrolyte LUMO coincides with the anode Fermi level, as has been widely
assumed in the literature.\cite{goodenough}  The LUMO lies above the organic
carbonate localized orbital and exhibits substantial statistical fluctuations;
however it may influence $e^-$ transfer kinetics if electron transfers
from the anode, through a substantial liquid layer, to FEC in the bulk
liquid electrolyte region, instead of towards FEC that diffuses near
the anode surface.

The above discussion adopts the solid state physics language often used in the
battery community.  It is important and of great interest to
reconcile our study with molecular electrochemistry terminology\cite{saveant}
less often featured in battery studies.  According to Marcus
theory,\cite{marcus} $e^-$ injection into FEC is accompanied with 
reorganization (free) energies ($\lambda$).  For FEC$^-$, $\lambda$ contains
a large intramolecular component, and is not solely due to ``outer shell''
solvation effects.  The ``polaronic shift'' of the HOMO of FEC$^-$, from
above the liquid conduction band edge if FEC$^-$ were flat, to within the
liquid gap due to FEC$^-$ geometry change and dielectric solvation, is a
non-trivial manifestation of this $\lambda$.  Quantitatively, the vertical
electron affinity should be at a value $\lambda$ above the molecular
reduction potential $\Phi$.\cite{adriaanse} Unfortuately, it is difficult
to compute $\lambda$ in simulation cells with electrodes which are electron
conductors.  For example, the liquid electrolyte LUMO at frozen liquid
geometry tend to reside above the Fermi level.  As a result, injecting
an $e^-$ to the system to calculate the vertical electron affinity immediately
populates the Fermi level of the electrode, not the the electrolyte, unless
constrained DFT methods are used in the electrolyte region.  In this sense,
the systems considered in this work may differ from electrodes with a
significant band gap, like TiO$_2$.\cite{cheng2014}  Thus all AIMD simulations
in this work report adiabatic free energy changes and redox potentials, not
vertical excitations that include $\lambda$.  To estimate $\lambda$, we have
applied a localized basis set and dielectric continuum approach similar to
Ref.~\onlinecite{cpl}.  We find that the total $\lambda$ for FEC is 3.2~eV.
This is somewhat larger than that of the structurally similar EC.\cite{cpl}
Our predicted $\lambda$ can be compared with future optical measurements in
organic solvents analogous to those in aqueous media.\cite{adriaanse}
Incidentally, the wide separation between the localized excess $e^-$ orbital
in FEC$^-$ and the liquid electrolyte HOMO minimizes hybridization between
the localized state and solvent orbitals, which has been shown to be
important for accurate DFT treatment of anions.\cite{adriaanse,galli}

\subsection{PF$_6^-$ Electrochemical Reduction}

Finally, this potential-calibration technique is used to investigate possible
PF$_6^-$ reductive decompostion on anode surfaces as ${\cal V}$ varies.  A
recent AIMD study has predicted rapid reductive decomposition of PF$_6^-$ into
LiF and PF$_n^{q-}$ fragments, $n\leq$4, at liquid EC/graphite edge
interfaces.\cite{kent}  Seemingly consistent with this prediction, simple
cluster-based calculations suggests that the LiPF$_6$ ``molecule'' has
a reduction potential of $1.46$~V (S.I.), far above LiC$_6$ potentials during
battery charging.  However, such a PF$_6^-$ reduction
signature has not been observed in cyclic voltametry.  Experimentally, the
SEI formed in LiPF$_6$-based electrolytes is known to contain LiF from
PF$_6^-$ breakdown, but it has been widely accepted that
PF$_6^-$ decomposes thermally or due to reaction with trace water over a
period of hours,\cite{kostecki_pf6,plak_pf6} not electrochemically in seconds.

\begin{figure}
\centerline{\hbox{ (a) \hspace*{0.02in} \epsfxsize=1.80in \epsfbox{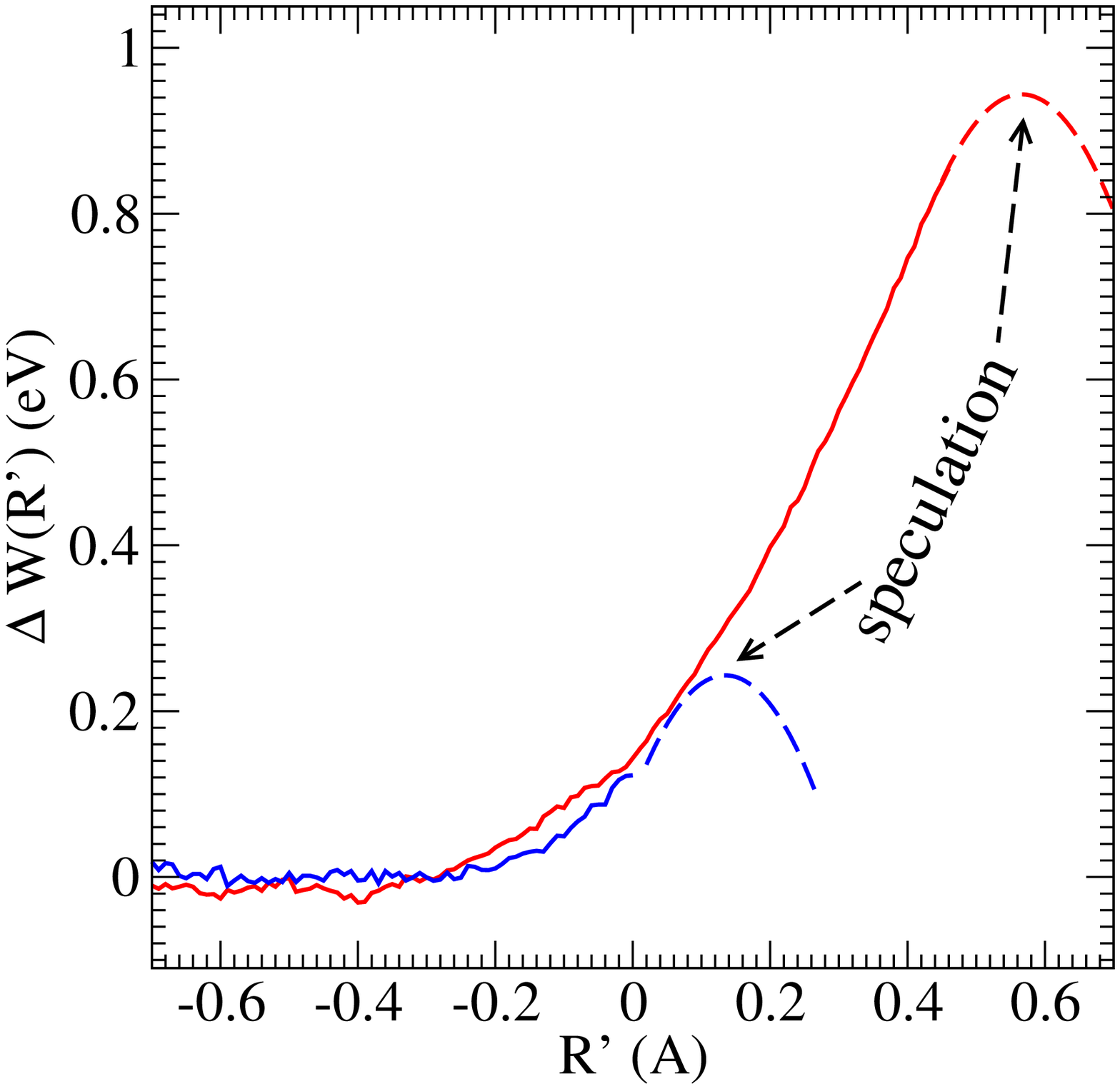} }
            \hbox{ (b) \epsfxsize=1.80in \epsfbox{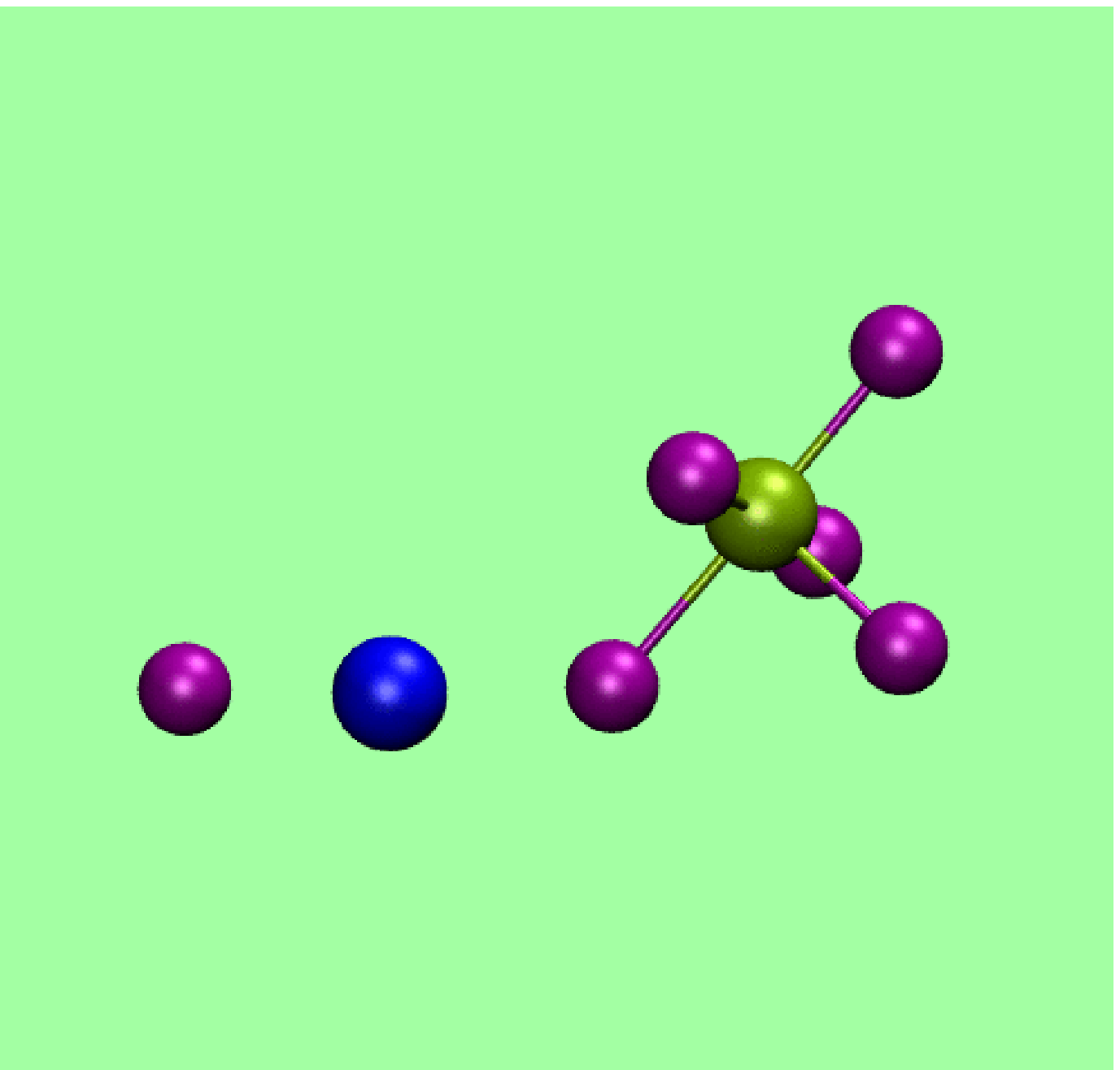} }}
\centerline{\hbox{ (c) \epsfxsize=1.80in \epsfbox{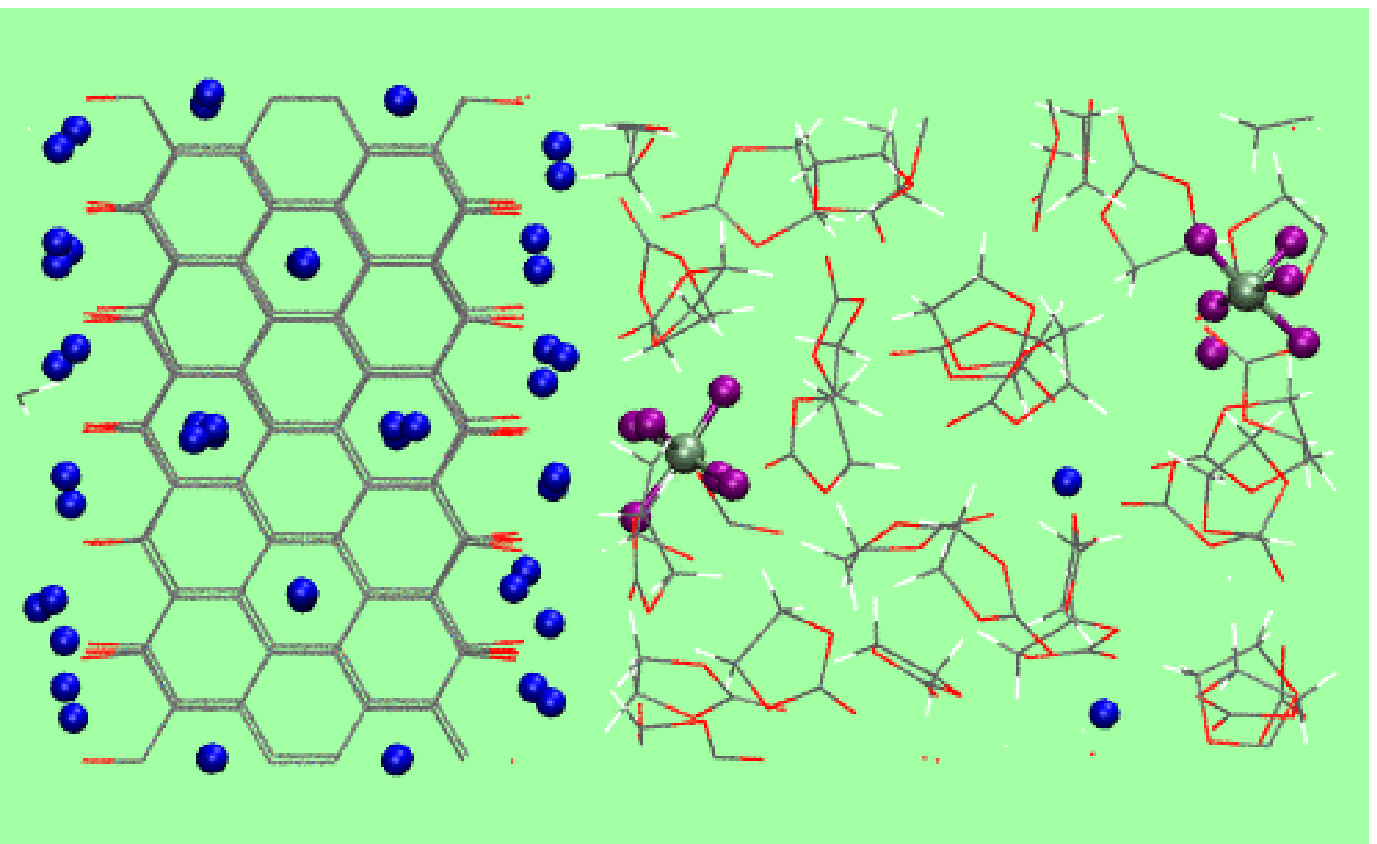} }
            \hbox{ (d) \epsfxsize=1.80in \epsfbox{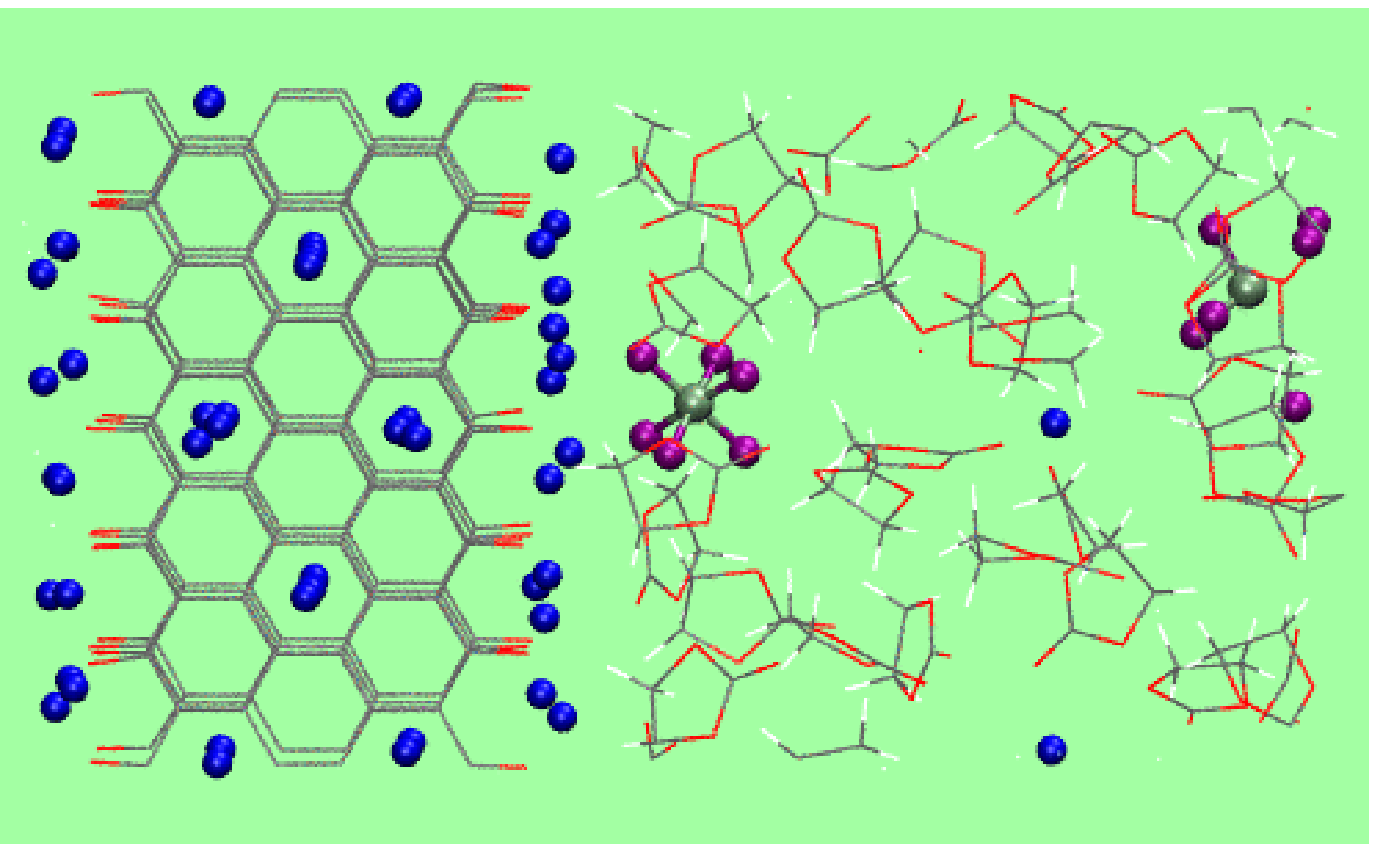} }}
\caption[]
{\label{fig4} \noindent
(a) Potentials-of-mean-force (PMF) of PF$_6$ electrochemical decomposition, as
functions $R'$ at estimated 0.56~V (red,
$n_{\rm Li}$=0.583) and $-$0.21~V (blue, $n_{\rm Li=0.750}$), respectively.  
The solid line portions depict actual data.  The barriers must reside in
the dashed line regions because PMF calculations there lead to spontaneous
PF$_6^-$ decomposition in picosecond timescale.  (b) Reduced
Li$^+$:F$^-$:PF$_5^-$ fragment from static, localized basis-set calculations.
(c)\& (d): AIMD snapshots in PMF trajectories at $V$=0.56~V, $R'$$\sim$$-0.21$
and $R'$$\sim$$+0.5$\AA, respectively.  The rightmost PF$_6^-$ has decomposed
in panel (d); recall that periodic boundary conditions are used.
}
\end{figure}

To reconcile these observations, we note that, unlike FEC reduction, $e^-$
transfer to PF$_6^-$ occurs in concert with P-F bond breaking.  This is
reminiscent of alkyl halide reduction,\cite{saveant} which is clearly voltage
dependent.  Previous PF$_6^-$ modeling work\cite{kent} has not specified its
anode potential.  In Fig.~\ref{fig4}, we apply AIMD potential-of-mean-force
($\Delta W(R)$) techniques to estimate the free energy barrier ($\Delta G^*$)
of PF$_6^-$ decomposition at two different potentials.

The trajectories include a PF$_6^-$ pre-equilibrated at each graphite edge,
charge-balanced by Li$^+$ in the electrolyte.  The 3-atom reaction coordinate is
$R'$=$|{\bf R_{\rm P}}-{\bf R}_{\rm F}|$$-$$|{\bf R_{\rm Li}}-{\bf R_{\rm F}}|$;
P, F, and Li atoms are chosen such that they are in position to react at
$t$=0~ps.  Increasing $R'$ is correlated with F$^-$ transfer from P to an
edge Li.  Umbrella sampling potentials of the form $A(R'-R_o)^2/2$ are
enforced, where $B$=4~to~10~eV and $R_o$ span the range between the reactant
and the transition state.  See the S.I.~for details.

Figure~\ref{fig4}a shows that, at $\cal V$=0.56~V ($n_{\rm Li}$=0.583),
$\Delta G^*$ is at least 0.9~eV.  As soon as one P-F bond breaks completely,
$e^-$ is transferred, other F$^-$'s detach from the P-atom
spontaneously (Fig.~\ref{fig4}d), and these irreversible steps render a
quasi-equilibrium sampling of $\Delta W(R)$ in the barrier top region
impossible.  In contrast, at ${\cal V}$=$-0.21$~V ($n_{\rm Li}$=0.75), the
barrier appears not much higher than 0.2~eV.  Fig.~\ref{fig4} indicates
that PF$_6^-$ electrochemical reduction may occur during the initial stage
of SEI formation if the anode is at sufficiently low potentials, although
this process faces competition from solvent reductive decomposition.

\section*{Conclusions}

In conclusion, we have calibrated the anode potential
(${\cal V}(\sigma=0,n_{\rm Li})$) of lithium intercalated graphite edge planes
at zero surface electronic charge ($\sigma$=0) as a function of the edge
Li content ($n_{\rm Li}$) by computing the free energy of Li$^+$ transfer
between electrode and liquid electrolyte.  The estimated ${\cal V}$ is
shown to be reasonable by correlating with observed electron transfer from
reduced fluoroethylene carbonate (FEC$^-$) radical anions inserted into the
liquid region.  Electrochemical reduction of PF$_6^-$ at the pristine edge
plane is shown to be viable at low potentials and to exhibit
potential-dependent kinetics.  This reduction pathway may need to be
considered during SEI formation, in addition to thermal/impurity water-induced
PF$_6^-$ decomposition routes widely accepted in the literature.  In the
future, optimization of the free energy with respect to all surface
parameters ($N_{\rm Li}$ and $\sigma$) will be performed, and our method
will be used to study the dynamics of Li$^+$ insertion into
passivated anodes as a function of the applied potential.

\section*{Acknowledgement}

We thank Kevin Zavadil and Jun Cheng for interesting discussions.  This work
was supported by Nanostructures for Electrical Energy Storage (NEES), an
Energy Frontier Research Center funded by the U.S.~Department of Energy,
Office of Science, Office of Basic Energy Sciences under Award Number
DESC0001160.  Sandia National Laboratories is a multiprogram laboratory
managed and operated by Sandia Corporation, a wholly owned subsidiary of
Lockheed Martin Corporation, for the U.S.~Deparment of Energy's National
Nuclear Security Administration under contract DE-AC04-94AL85000.

$\dagger$ Electronic supporting Information available.  See DOI:TBA.


\begin{references}
\bibitem{conceit}
We have used LiC$_6$ inside the electrode throughout, instead of pure graphite
at high voltage and LiC$_6$ at low.  The rationale is that the electrolyte
does not see the electrode interior. See Ref.~\onlinecite{ec} for a discussion.
\bibitem{note3}
Organic carbonates have been shown to exhibit higher second reduction
potentials than first in high dielectric media, especially if they have
broken bonds.\cite{cpl}
\bibitem{note4}
Although Kohn-Sham orbitals are not physical quantities in general and the
above discussion is therefore qualitative, at the onset of FEC reduction,
when the occupied FEC$^-$ HOMO orbital is at the Fermi level, the
change of total electronic energy as a function of the partial occupancy
of this orbital should indeed correspond to the orbital energy.  In that
sense Fig.~3 is well defined at the onset of FEC reduction.
\end{references}
\end{document}